\documentstyle[11pt]{article}

\small
\begin{document}

\title{
{\bf Effective Lagrangian with vector mesons : \\
Linear response theory }
}
\author{ {\bf Abdellatif Abada}
\thanks {On leave from :
 Division de Physique Th\'eorique, Institut de Physique Nucl\'eaire,
F-91406 Orsay }
\thanks  {email : "ABADA@frcpn11.in2p3.fr"}
\\
{\normalsize
Laboratoire de Physique Nucl\'eaire, CNRS/IN2P3 Universit\'e de Nantes, }\\
{\normalsize	2, Rue de la Houssini\`ere, 44072 Nantes Cedex 03, France }
}

\date{}
\maketitle
\begin{abstract}
The soliton breathing mode is investigated in the framework of linear response
theory within a Skyrme model vector meson stabilized.
The effective Lagrangian considered includes the $\rho$ (introduced following
the standard prescription of nonlinear chiral symmetry) and the $\omega$
mesons.
The monopole response function is found to have a pronounced peak
which is identified to the $P11$ (Roper) resonance.
The results are compared to those obtained within the local approximation.
\end{abstract}


\vskip 1cm
LPN 93-12

IPNO/TH 93-47 \hfill{October 1993}

\vskip 1cm
{\it Submitted to Phys. Lett. B }

\newpage
In a recent paper \cite{AM} we have considered an extended Skyrme model which
includes fourth and sixth-order terms and have investigated the soliton
breathing mode
in the framework of the linear response theory. We have found that the monopole
response function exhibits an unbound sharp peak which we have identified to
the Roper resonance $N$(1440). Since the fourth and sixth-order terms can be
derived from a local approximation of an effective model with $\rho$
\cite{Ik} and $\omega$ mesons \cite{Jac,Ad2} respectively, it seems therefore
interesting to explore the soliton breathing mode, with the same method as in
\cite{AM} (first proposed in Ref. \cite{AV}), within an effective Lagrangian
which incorporates these two mesons {\it explicitely}.
This is the purpose of the present paper.

\vskip .5cm
\hskip .5cm {\Large {\bf 1.}}
{}~Our starting point is the following Lagrangian density:
\begin{equation} \label{la}
{\cal L} = {\displaystyle \frac {f_{\pi}^2 }{4}
{}~\hbox {Tr} ~(\partial_{\mu} U \partial^{\mu} U^{\dagger})
+ {\cal L}_{\pi \rho} + {\cal L}_{\pi \omega}
+ \frac{1}{4} f_{\pi}^2 m_{\pi}^2 \hbox{Tr} (U + U^{\dagger} - 2 ) }
\end{equation}
where $U$ is an SU(2) matrix parametrized by the pion fields
$\pi_a$, normalized to the pion decay constant $f_{\pi}$ :
\begin{equation} \label{pion}
U = {\displaystyle \exp[~i \vec {\pi}. \vec {\tau} / f_{\pi} ~] }~,
\end{equation}
the $\tau_{a}$'s being the usual Pauli matrices.
The first term in Eq. (\ref{la}) corresponds to the well known nonlinear
$\sigma$ model. The
second one corresponds to the coupling between the isospin one
$\rho$-meson and the pion field. The $\rho$ vector meson is frenquently
introduced in the chiral Lagrangian as hidden gauge
particle \cite{KO} or as a massive Yang-Mills field \cite{FJ}. In both cases
the soliton was shown to be {\it unstable} and consequently one can not
describe
the excited states, in particular the breathing mode. In a recent paper
\cite{AKM} it was proven that the soliton is stable with respect to the
breathing fluctuations if the $\rho$-meson
is introduced in the chiral Lagrangian following the standard prescription of
nonlinear chiral symmetry for massive particles \cite{CW}.
In Ref. \cite{AKM} however,
the $\rho$-meson is described in terms of antisymmetric tensor field as
proposed by the authors of Refs. \cite{GL,EG}. It has been shown in Ref.
\cite{Ka} that this description is canonically equivalent to a vector field
formulation provided the Skyrme term \cite{Sk} is added to the latter
in order to keep the corresponding Hamiltonian bounded.
In this work we make use of this last formulation to introduce the
$\rho$-meson. It yields:
\begin{equation} \label{rho}
{\cal L}_{\pi \rho} = {\displaystyle
\frac{M_{\rho}^2}{2} \hbox {Tr} (V_{\mu}V^{\mu}) - \frac{1}{4} \hbox {Tr}
\left \{~(~\nabla_{\mu} V_{\nu} - \nabla_{\nu} V_{\mu} +
i \frac{g_V}{\sqrt 2} [u_{\mu},u_{\nu}]~)^2 ~\right \}	}
\end{equation}
where $V_{\mu}$ is the $\rho$-meson field and
\begin{equation} \label{def}
{\displaystyle
\nabla_{\mu} = \partial_{\mu}  + [\Gamma_{\mu},~ ], ~
\Gamma_{\mu} = \frac{1}{2}[u^{\dagger},\partial_\mu u], ~
u = U^{\frac{1}{2}}, ~
u_{\mu} = i u^{\dagger} \partial_{\mu} U  u^\dagger	}~.
\end{equation}
The two constants appearing in Eq. (\ref{rho}) are the $\rho$-meson mass
$M_{\rho}$ and the dimensionless parameter $g_V$ which can be related to the
$\rho \to \pi \pi$ decay width.

The third term in Eq. (\ref{la}) corresponds to $\omega$-meson exchange
\cite{Ad2}. It reads:
\begin{equation} \label{omega}
{\cal L}_{\pi \omega} = {\displaystyle
-\frac{1}{4} (~\partial_{\mu} \omega_{\nu} - \partial_{\nu} \omega_{\mu}~)^2
+ \frac{1}{2} m_{\omega}^2 \omega_{\mu} \omega^{\mu} +
\beta_{\omega} \omega_{\mu} B^{\mu} }
\end{equation}
where $\omega_{\mu}$ is the omega field and $B^{\mu}$ the baryon current
\cite{Sk,Wi}
\begin{equation} \label{Bmu}
B^{\mu} = {\displaystyle \frac{1}{24 \pi^2}
\epsilon ^{\mu \nu \alpha \beta} \hbox {Tr} \left \{~
(\partial_{\nu} U) U^{\dagger} (\partial_{\alpha} U) U^{\dagger}
(\partial_{\beta} U) U^{\dagger}
{}~\right \} } .\end{equation}
The two new constants appearing in Eq. (\ref{omega}) are the $\omega$-meson
mass $m_{\omega}$ and the parameter $\beta_{\omega}$ which can be related
to the $\omega \to \pi \gamma $ width \cite{La}.
Finally, the last term in the Lagrangian (\ref{la}) implements a small explicit
breaking of chiral symmetry, $m_{\pi}$ being the pion mass.

\vskip .5cm
\hskip .5cm {\Large {\bf 2.}}
{}~In order to work with true degrees of freedom and eliminate the constrained
fields $V_0$ and $\omega_0$ (the corresponding canonical momenta conjugate are
equal to zero), it is suitable to consider the Hamiltonian density.
Let $\vec \Phi, P^i$ and $p^i$ being the canonical momenta conjugate to the
pion field $\vec \pi$, the $\rho$ field and the $\omega$ field
respectively. We make the hedgehog ansatz for the pion and the
most general spherical ansatz for the $\rho$ and the $\omega$ fields:
\begin{equation} \label{ansatz} \begin{array}{ll}
&{\displaystyle
\vec \pi = f_{\pi} F ~\hat{{\bf r}},~~
V_i = f_{\pi} \left \{ v_1(\tau_i - (\vec \tau. \hat{{\bf r}}) \hat {r}_i)
+ v_2 (\vec \tau.\hat{{\bf r}}) \hat {r}_i
- v_3 (\vec \tau \times \hat{{\bf r}})_i \right \},~~
\omega_i = f_{\pi} \omega ~\hat {r}_i, }\\
&{\displaystyle
\vec \Phi = \frac {f_{\pi}^2}{r} \phi ~\hat{{\bf r}},~~
P_i = \frac {f_{\pi}}{r} \left \{ \gamma_1(\tau_i - (\vec \tau. \hat{{\bf r}})
\hat {r}_i) + \gamma_2 (\vec \tau.\hat{{\bf r}}) \hat {r}_i
- \gamma_3 (\vec \tau \times \hat{{\bf r}})_i \right \},~~
p_i = \frac {f_{\pi}}{r} p ~\hat {r}_i ~. }
\end{array} \end{equation}
The energy functional expressed in terms of the time-dependent radial
functions $F, \phi, v_i, \gamma_i, \omega, p$ reads:
\begin{equation} \label{energy} \begin{array}{ll}
E = &{\displaystyle 4 \pi f_{\pi}^2 \int_0^{\infty} \hbox {d}r \{~
\frac{1}{2} \gamma_1^2 + \frac{1}{4} \gamma_2^2 + \frac{1}{2} \gamma_3^2 +
M_{\rho}^2r^2(2v_1^2+v_2^2) + \frac{1}{2} m_{\omega}^2r^2 \omega^2 + } \\
&{\displaystyle
\frac{1}{2}(\phi+2\bar g\frac{s}{r}\gamma_3 -\bar{\beta}\frac{s^2}{r}\omega)^2
+ 2((rv_1)'-v_2c)^2 + \frac{1}{4M_{\rho}^2r^2}(2\gamma_1c-(\gamma_2r)')^2 + }\\
&{\displaystyle \frac{1}{2} r^2F'^2 + s^2 + 2M_{\rho}^2r^2v_3^2 +
\frac{1}{2}p^2  + m_{\pi}^2r^2(1-c) +}\\
&{\displaystyle (2v_3c+\bar g \frac{s^2}{r})^2 + 2(\bar gsF' +(rv_3)')^2 +
\frac{1}{2m_{\omega}^2r^2}(\bar {\beta}s^2F'+(rp)')^2 } ~\}
\end{array} \end{equation}
with $\bar g = \sqrt 2 g_V/f_{\pi}$,
$\bar {\beta} = \beta_{\omega}/2\pi^2f_{\pi}$, $s=\sin F$ and $c=\cos F$.
Primes
indicate radial differentiations.

By solving the static Hamilton equations, i.e.,
$$
\dot F = \dot \phi = \dot v_i = \dot \gamma_i = \dot \omega = \dot p = 0
$$
where dots indicate partial time differenciations,
we find that the only nonsingular
solution which extremizes $E$ is the one which corresponds to
\begin{equation} \label{sol1}
\phi(r) = v_1 = v_2 = \gamma_1 = \gamma_2 = \gamma_3 = \omega = 0~.
\end{equation}
The set of the three other static Hamilton equations corresponding to
$F,v_3$ and $p$ reads:
\begin{equation} \label{sol2} \begin{array}{ll}
&{\displaystyle
r(rF)''+ 4\bar g \frac{s}{r}(M_{\rho}^2r^2+s^2)v_3+2sc(4v_3^2-1)+\bar
{\beta} \frac {s^2}{r}p - m_{\pi}^2r^2s = 0~,}\\
&{\displaystyle
r(rv_3)''-(2c^2+M_{\rho}^2r^2)v_3 -\bar g(\frac{c s^2}{r} + rc'') = 0~,}\\
&{\displaystyle
r^2p''-(2+m_{\omega}^2r^2)p +r^3\bar {\beta} (\frac{s^2}{r^2}F')' =0 ~.}
\end{array} \end{equation}
We solve numerically the equations (\ref{sol2}) with the boundary
conditions $F(0) = \pi, ~F(\infty) = 0$ in order to find a winding
number one solution. In Fig. 1 we plot that solution for the following
set of physical parameters (fixed by fitting to the low energy meson
observables) :
\begin{equation} \label{parameter}
\begin{array}{ll}
f_{\pi} = 93, ~M_{\rho}= 769, ~m_{\omega}=782, ~m_{\pi}=139.5 ~~~(\hbox{MeV})
\\
\beta_{\omega} = 9.3, g_V = 0.09 ~.
\end{array}
\end{equation}
The classical soliton mass we find is $M=1515$ MeV, and is lower than
the mass which corresponds to the local approximation (1714 MeV) that one
finds if the masses $M_{\rho}$ and $m_{\omega}$ and the coupling constant
$\beta_{\omega}$ are increased to infinity,
keeping $g_V$ and the ratio $\beta_{\omega}/m_{\omega}$ finites.
We also plot in Fig. 1 the local approximation's chiral function for reference.

\vskip .5cm
\hskip .5cm {\Large {\bf 3.}}
In order to investigate the soliton breathing mode within the model (\ref{la})
we consider the linearized time-dependent Hamilton equations in
presence of an external infinitesimal monopole field with a frequency $\Omega$.
This can be done by adding the following perturbation term \cite{AM,AV}
\begin{equation} \label{pertur}
\delta E_{\hbox {int}} = \epsilon f_{\pi}^3 \langle r^2\rangle
\sin(\Omega t) \exp(\eta t) \end{equation}
to the energy functional (\ref{energy}). In Eq. (\ref{pertur}),
$\langle r^2\rangle$ is the isoscalar mean square radius \cite{Ad1}
and $\eta$ a vanishingly small positive number.

Because the term (\ref{pertur}) is weak it introduces only small changes of the
classical Hamilton solution. We thus can use the linear approximation
and look for a solution in the form
$$ \begin{array}{ll}
&{\displaystyle F(r,t) = F(r) +  \epsilon [~\delta F(r)
e^{i(\Omega - i\eta)t} - c.c~]/2i ~,~
\phi(r,t) = \phi(r) +  \epsilon [~\delta \phi(r)
e^{i(\Omega - i\eta)t} - c.c~]/2i~, }\\
&{\displaystyle \omega (r,t) = \omega (r) +  \epsilon [~\delta
\omega (r)  e^{i(\Omega - i\eta)t} - c.c~]/2i ~,~
p(r,t) = p(r) +  \epsilon [~\delta
p(r)  e^{i(\Omega - i\eta)t} - c.c~]/2i~, }\\
&{\displaystyle v_i(r,t) = v_i(r) +  \epsilon [~\delta v_i(r)
e^{i(\Omega - i\eta)t} - c.c~]/2i ~,~
\gamma_i (r,t) = \gamma_i (r) +  \epsilon [~\delta
\gamma_i (r)  e^{i(\Omega - i\eta)t} - c.c~]/2i~, } ~~
\end{array}
$$
with the boundary conditions that all the fluctuations and their time
derivatives are equal to zero at $t=-\infty$.
We are not going to write all the ten linearized equations but it is
straightforward to show from Eq. (\ref{energy}) that the equations
corresponding to the fluctuations $\delta \phi, \delta \omega, \delta
\gamma_i, \delta v_1$ and $\delta v_2$ decouple from those which correspond to
$\delta F$, $\delta v_3$ and $\delta p$, and consequently do not contribute to
the breathing mode. The equations corresponding to the breathing fluctuations
$\delta F$, $\delta v_3$ and $\delta p$ read :
\begin{equation} \label{linea}
\left [~~ (\Omega-i\eta)^2 ~-~
\pmatrix{ A_{11} & A_{12} & A_{13}
\cr A_{21} & A_{22} & 0
\cr A_{31} & 0 & A_{33} \cr} ~~\right]
{\displaystyle
\pmatrix{ \delta F \cr \delta \psi \cr \delta \xi \cr} ~=~
\frac {f_{\pi}}{\pi^2} \frac{s^2}{r} \pmatrix{ 1 \cr 0 \cr 0 \cr} }
\end{equation}
where $\delta \psi$ and $\delta \xi$ are combinations of $\delta v_3$ and
$\delta p$, namely
$$ {\displaystyle
\delta \psi = \delta v_3 + \bar g \frac {s}{r} \delta F ~,~
\delta \xi = \delta p + \bar {\beta} \frac {s^2}{r} \delta F }~.
$$
The matrix elements $A_{ij}$ of the operator ${\cal A}$ appearing in Eq.
(\ref{linea}) are themselves operators and read as:
\begin{equation} \label{operators} \begin{array} {ll}
\begin{array}{rr}
{\displaystyle
A_{11} = -\frac {1}{r} \frac{d^2}{dr^2} r + \frac{4}{r^2} [~
(2v_3^2-\frac{1}{2})(s^2-c^2) +
\frac{\bar g}{r} s^2(\frac{\bar g}{r}s^2 +v_3 c) +
\frac{\bar g}{r} M_{\rho}^2r^2 (\frac{\bar g}{r}s^2 - v_3 c) }\\
{\displaystyle
- \frac{\bar {\beta}}{4r}s (2cp- \frac {\bar {\beta}}{r}s^3) +
\frac{1}{4} m_{\pi}^2r^2c  ~]
}\end{array}
\\
{\displaystyle
A_{12} = -\frac{4s}{r^2} [~\frac{\bar g}{r}(s^2+M_{\rho}^2r^2) + 4v_3c ~]
}\\
{\displaystyle
A_{13} = -\bar {\beta} \frac{s^2}{r^3}
}\\
{\displaystyle
A_{21} = -\frac{s}{r^2} [~\frac{\bar g}{r}(s^2+M_{\rho}^2r^2) + 4v_3c ~]
}\\
{\displaystyle
A_{22} = -\frac {1}{r} \frac{d^2}{dr^2} r + \frac{1}{r^2}
(~2c^2+M_{\rho}^2r^2)
}\\
{\displaystyle
A_{31} = -m_{\omega}^2 \bar {\beta} \frac{s^2}{r}
}\\
{\displaystyle
A_{33} = - \frac{d^2}{dr^2} + \frac{1}{r^2}
(m_{\omega}^2r^2 +2)
}
{}~~.\end{array} \end{equation}
We recall that $s$ means $\sin (F)$, $c$ stands for $\cos (F)$ and $F$ is the
static solution for the pion field.

The monopole response function is determined from the
evolution of the isoscalar mean square radius of the soliton with respect to
the frequency $\Omega$. The mean square radius is given by \cite{Ad1}
$${\displaystyle
\langle r^2 \rangle (t) ~=~ \int \hbox{d}^3 r B^0(r,t) r^2 ~=~
-\frac {2}{\pi }
\int_0^\infty  \hbox{d} r ~r^2 \sin^2(F) F' }$$
where $B^0(r,t)$ is the time-component of the baryon current (\ref{Bmu}).
Up to first order in $\epsilon$, $\langle r^2\rangle$ reads
$$ {\displaystyle
\langle r^2 \rangle (t) = \langle r^2 \rangle_0 +\frac {\epsilon}{2i}
[~f( \Omega) e^{i(\Omega -i\eta) t} - c.c ~]
}$$
where $f$ is the linear response function
\begin{equation} \label{response}
f( \Omega) = {\displaystyle
\frac{4}{\pi} \int_0^\infty \hbox{d} r ~r \sin^2(F) \delta F(r) } .
\end{equation}
The spectral representation of the response function $f$ can be extracted by
using equations (\ref{linea}). It reads
\begin{equation} \label{spectral}
{\displaystyle
f( \Omega) = \frac {1}{\pi} \sum_n
\frac {| \langle \chi \vert \chi_n\rangle |^2 }
{ (\Omega - i\eta)^2 - \lambda_n^2 } } ~,\end{equation}
where the limit $\eta \to 0^+ $ is, as usual, implicit,
and corresponds to the boundary conditions specified above.
In Eq.(\ref{spectral}) the state $\chi$ is defined by
\begin{equation} \label{chi}
{\displaystyle
\langle \chi\vert r \rangle =
\frac {2}{\pi f_{\pi}} \frac{s^2}{r} (1 ~,~ 0 ~,~ 0 ) }
\end{equation}
and the $\chi_n$ are the eigenstates of the operator ${\cal A}$ (see Eqs.
(\ref{linea}),(\ref{operators})), with the eigenvalues $\lambda^2_n$,
normalized according to
$${\displaystyle
\langle \chi_n|\chi_m\rangle   =
\int_0^{\infty} f_{\pi}^3 r^2 \hbox {d} r~
\chi_n^+(r) ~\chi_m(r) = \delta_{nm} } ~~.
$$

Since the distribution of collective strength is directly related to
the imaginary part of the linear response function, we should
consider this quantity.
We display in Fig. 2 the imaginary part of the function
(\ref{spectral}) with respect to the frequency $\Omega$ for the parameters
given in Eq. (\ref{parameter}). We find that it
exhibits a pronounced peak at 400 MeV which we identify with the excitation
energy of the Roper resonance. In Ref. \cite{AM} we have found nearly the same
value within the local approximation to the Lagrangian (\ref{la}) except for
the detail that the dimensionless Skyrme term parameter $e$ \cite{Sk} taken
in \cite{AM} is equal to 7.1 \cite{Ri,Mo}, while the value of $e$ which
corresponds to the parameters (\ref{parameter}) we use in this work is
$e=1/2g_V=5.57$. We show in Fig. 2 the response function which corresponds to
the local approximation \cite{AM} taking $e=5.57$. One sees from this figure
that there
is an improvement over the local case, although a slight one ($\sim 40$ MeV)
but in the good direction. A more realistic improvement consists of extending
the model (\ref{la}) to include other mesons such
as the vector meson-$A_1$ and the scalar meson-$\epsilon$
despite of the increasing complexity implied.

\vskip .5cm
\hskip .5cm {\Large {\bf 4.}}
To summarize, we have investigated the soliton breathing mode within an
effective Lagrangian with finite mass mesons (the $\rho$- and the
$\omega$-mesons) in the framework of linear response theory. Our results for
the location of the Roper resonance show an improvement
over the local approximation where these two mesons are taken to be infinitely
heavy.

It is quite well established (see for example \cite{La}) that for the
description of static nucleon properties, finite mass meson Lagrangians are
more accurate that their local residues, such as the Skyrme model. As this work
suggests, this feature is also true for the description of dynamical
observables such as the breathing mode.

Finally, the difficulty to find the $P11$ resonance $N(1440)$ in analysis based
on phase shifts method \cite{SW} and the encouraging results obtained in
this work and in Refs. \cite{AM,AV} lead us to think that the linear response
theory is the most adapted method to describe the Roper resonance within
effective Lagrangians.

We are grateful to J. Aichelin, D. Kalafatis, B. Moussallam and
D. Vautherin for a critical reading of the manuscript.

\newpage
{\Large {\bf Figure captions } }
\vskip .5cm

{\bf FIG. 1.} The solution of Eq. (\ref{sol2}) minimizing the energy functional
(\ref{energy}). The full line displays the static function $F(r)$ in
this model, while the dotted one corresponds to the local approximation of
(\ref{la}). In the dashed and dashed-dotted lines we display the $\rho$ and the
$\omega$ degrees of freedom respectively. Namely $v_3(r)$ and $p(r)$.

{\bf FIG. 2.}  Imaginary part of the response function $f $~(fm$^2)$
versus the energy $ \Omega $ in MeV. The full line corresponds to the
Lagrangian (\ref{la}) with the parameters given in Eq. (\ref{parameter})
while the dotted line corresponds to the local approximation of (\ref{la}) with
$f_{\pi} =$ 93 MeV, $\beta_{\omega} =$ 9.3 and $e =1/2g_V = 5.57$ (c.f. Ref.
\cite{AM}).

\end{document}